\documentstyle[aps,prl,multicol,color]{revtex}
\input {epsf.sty}
\begin{document}
\draft
\vspace{-11pt}
\begin{multicols}{2}
{\bf \noindent Comment on "Magnetic Field Independence of 
the Spin Gap in YBa$_2$Cu$_3$O$_{7-\delta}$"}
\\

The nature of the onset of superconductivity in high temperature
superconductors has been of considerable recent interest.  Does the
superconducting order parameter emerge from preformed pairs, or reflect the
opening of a spin-pseudogap or a pseudo-gap in the charge or pairing
channels?  Recent papers that relate to this subject are the measurement
of the magnetic field dependence of the $^{63}$Cu(2) NMR spin lattice relaxation in
YBCO by Gorny {\it et al.}\cite{Gorny99} and Mitrovi{\'c} {\it et al.}\cite{Mitrovic}
and the related experiments on the spin susceptibility taken from the oxygen
NMR shift of Bachman {\it et al.}\cite{bachman98b}.

\begin{figure}[h]
\centerline{\epsfxsize0.95\hsize\epsffile{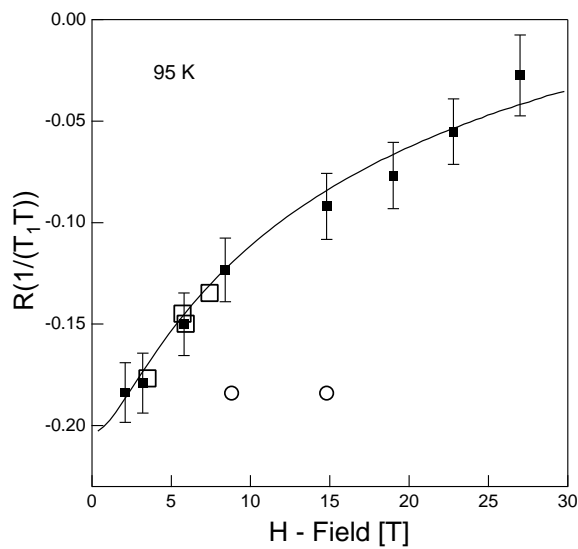}}
\begin{minipage}{0.95\hsize}
\caption[]{\label{Fig1}\small
$R(1/(T_1T))$ = 
$((T_1T)^{-1}_{meas.} - (T_1T)^{-1}_{n}) / (T_1T)^{-1}_{n}$ of $^{63}$Cu(2) rate as a function of
magnetic field at 95 K, the closed squares from Mitrovi{\'c} {\it et al.} \cite{Mitrovic}.
$(T_1T)^{-1}_{n}$ is found
from a fit to high temperature behavior and is given by 1648 s$^{-1}$/(103 K +
$T$).   The solid curve is  calculated pairing fluctuation contribution at 95 K \cite{Eschrig98}. 
The open  circles are the measurements of Gorny {\it et al.}\cite{Gorny99}. The open squares are
from  Song (3.5 T) \cite{song91}, Auler {\it et al.} (5.7 T) \cite{auler}, Carretta
{\it et al.}  (5.9 T) \cite{carretta96},  and  Hammel {\it et al.} (7.4 T) \cite{hammel89}.
}
\end{minipage}
\end{figure}

In the recent letter 
of Gorny {\it et al.}\cite{Gorny99} it was reported that the spin-lattice 
relaxation rate in YBCO$_{7-\delta}$  is magnetic field  independent and
indicative of the opening of a spin-pseudogap.  These authors point out
that their results are inconsistent with prior reports \cite{Mitrovic,borsa92,carretta96}
which concluded that there is a small but significant magnetic field dependence to $1/T_1T$
near $T_c$ in optimally doped material with interpretation in terms of pairing 
fluctuations \cite{Mitrovic,carretta96} (see $\mbox{Fig. \ref{Fig1}}$). 
Gorny {\it et al.}\cite{Gorny99} suggest that previous experiments might be in error since
special care is required in the measurement of the
$^{63}$Cu(2) NMR rate in order to avoid what they refer to as background
effects.  These effects, and how to deal with them in aligned powder
samples, have been known and practiced for some time \cite{song91,martindale}. The
particular approach of Gorny {\it et al.}, measuring a satellite resonance, appears to be
quite effective,  but not unique.

	In this comment we propose a different explanation for the discrepancy between the reports.
Quite simply, the sample of Gorny {\it et al.} is not optimally doped.  
First,  there is  evidence for this in the distribution of electric field
gradients apparent in the spectrum published by Gorny {\it et al.}, being 
twice as broad (FWHM $\approx$ 400kHz) as is expected for high-quality, 
optimally doped material (FWHM $<$ 200kHz).  In addition, field independence for the NMR in
slightly underdoped materials has been noted previously by Auler {\it et al.}\cite{auler} who
have shown that NQR and NMR (H = 12 T) have the same rates for this resonance near $T_c$. 
Finally, Song\cite{song91} $(\Delta \delta \approx 0.04)$ and Auler {\it et al.}\cite{auler}
$(\Delta
\delta \approx 0.02)$ have shown the effect of modest underdoping is to shift the maximum value
of
$1/T_1T$ to lower values and to higher temperatures compared with optimal doping. This explains
why the Gorny {\it et al.}\cite{Gorny99} measurements, open circles in $\mbox{Fig.\ref{Fig1}}$,
fall below others at 95 K. 

Results for optimally doped materials are remarkably consistent\cite{Mitrovic}, having a
magnetic field dependence shown in the figure for the data at a temperature of 95 K.  This field
dependence close to $T_c$ has been successfully interpreted in terms of d-wave superconducting
pairing fluctuations \cite{Mitrovic,Eschrig98}. In the underdoped
materials the fluctuation effects seem to be dominated by a  field-independent 
\noindent  
pseudogap,
developing rapidly near optimal doping \cite{auler},
 reported by Carretta {\it et
al.}\cite{carretta98}, Auler {\it et al.}\cite{auler}, and confirmed by  Gorny {\it et
al.}\cite{Gorny99}.
\\
\\
V. F. Mitrovi{\'c}, H. N. Bachman, W. P. Halperin,\\ M. Eschrig, J. A.  Sauls

\begin{verse}
Department of Physics and Astronomy, and \\ Science and Technology Center for \\
Superconductivity,
Northwestern University,\\ Evanston, Illinois 60208\\
\end{verse}
(PACS numbers: 74.25.Nf, 74.40.+k, 74.72.Bk) 

\vspace{-0.5cm}
\bibliographystyle{unsrt}

\vspace{0.5cm}
\end{multicols}
\end{document}